\documentclass[11pt,a4paper]{article}
\pdfoutput=1
\usepackage{jcappub}
\usepackage{amsmath,amssymb}
\usepackage{amsthm}
\usepackage{mathtools}
\usepackage{appendix}
\usepackage{comment}
\usepackage{graphicx}
\usepackage{grffile}
\usepackage{mathrsfs}
\usepackage{bm}
\usepackage{url}
\usepackage{xcolor}
\usepackage{soul}
\usepackage{algorithm}
\usepackage{algorithmic}
\usepackage{natbib}
\usepackage{physics}
\usepackage{braket}
\usepackage{siunitx}
\usepackage{acronym}
\usepackage{xspace}

\hypersetup{colorlinks=true
,urlcolor=DARKBLUE
,anchorcolor=DARKBLUE
,citecolor=DARKBLUE
,filecolor=DARKBLUE
,linkcolor=DARKBLUE
,menucolor=DARKBLUE
,linktocpage=true
,pdfproducer=medialab
,pdfa=true
}

\definecolor{MONZA}{HTML}{CF000F}
\definecolor{DARKBLUE}{HTML}{00008b}
\definecolor{DARKMAGENTA}{HTML}{8b008b}
\definecolor{DARKCYAN}{HTML}{00cfc0}

\newcommand{\Mpl}{M_\mathrm{Pl}}

\newcommand{\uG}{\mathrm{G}}

\newcommand{\bfk}{\mathbf{k}}

\newcommand{\calO}{\mathcal{O}}
\newcommand{\calP}{\mathcal{P}}
\newcommand{\calR}{\mathcal{R}}

\newcommand{\bfx}{\mathbf{x}}

\newcommand{\bae}[1]{\begin{align} #1 \end{align}}
\newcommand{\bce}[1]{\begin{cases} #1 \end{cases}}

\begin{document}
\title{Constant roll and non-Gaussian tail in light of logarithmic duality}

\author[a]{Ryoto Inui,}
\author[b]{Hayato Motohashi,}
\author[c,d,e]{Shi Pi,}
\author[a,f]{Yuichiro Tada,}
\author[g,a,e]{and Shuichiro Yokoyama}

\affiliation[a]{Department of Physics, Nagoya University, 
Furo-cho Chikusa-ku,
Nagoya 464-8602, Japan}
\affiliation[b]{Division of Liberal Arts, Kogakuin University, 2665-1 Nakano-machi, Hachioji, Tokyo, 192-0015, Japan}
\affiliation[c]{CAS Key Laboratory of Theoretical Physics, Institute of Theoretical Physics,
Chinese Academy of Sciences, Beijing 100190, China}
\affiliation[d]{Center for High Energy Physics, Peking University, Beijing 100871, China}
\affiliation[e]{Kavli Institute for the Physics and Mathematics of the Universe (WPI), UTIAS, The University of Tokyo, Kashiwa, Chiba 277-8583, Japan}
\affiliation[f]{Institute for Advanced Research, Nagoya University,
Furo-cho Chikusa-ku, 
Nagoya 464-8601, Japan}
\affiliation[g]{Kobayashi Maskawa Institute, Nagoya University, 
Chikusa, Aichi 464-8602, Japan}

\emailAdd{inui.ryoto.a3@s.mail.nagoya-u.ac.jp}
\emailAdd{motohashi@cc.kogakuin.ac.jp}
\emailAdd{shi.pi@itp.ac.cn}
\emailAdd{tada.yuichiro.y8@f.mail.nagoya-u.ac.jp}
\emailAdd{shu@kmi.nagoya-u.ac.jp}

\date{\today}
\abstract{
The curvature perturbation in a model of \ac{CR} inflation is interpreted in view of the logarithmic duality discovered in Ref.~\cite{Pi:2022ysn} according to the $\delta N$ formalism.
We confirm that the critical value $\beta\coloneqq\ddot{\varphi}/(H\dot{\varphi})=-3/2$ determining whether the \ac{CR} 
condition is stable or not is understood as the point at which the dual solutions, i.e., the attractor and non-attractor solutions of the field equation, are interchanged.
For the attractor-solution domination, the curvature perturbation in the \ac{CR} model is given by a simple logarithmic mapping of a Gaussian random field, which can realise both the exponential tail (i.e., the single exponential decay) and the Gumbel-distribution-like tail (i.e., the double exponential decay) of the probability density function, depending on the value of $\beta$.
Such a tail behaviour is important for, e.g., the estimation of the \ac{PBH} abundance. 
}
\maketitle

\acrodef{PDF}{probability density function}
\acrodef{PBH}{primordial black hole}
\acrodef{GW}{gravitational wave}
\acrodef{CMB}{cosmic microwave background}
\acrodef{SR}{slow-roll}
\acrodef{USR}{ultra-slow-roll}
\acrodef{CR}{constant-roll}

\acresetall
\section{Introduction}

The \ac{SR} inflation scenario is a leading paradigm of the early universe.
It naturally explains a long-lasting inflationary phase and also generates the almost scale-invariant power spectrum of the primordial scalar perturbation confirmed by cosmological observations such as measurements of the \ac{CMB} anisotropies~\cite{Planck:2018jri}.
The scale-invariant spectrum can be realised also in the other limiting case: \ac{USR} inflation~\cite{Tsamis:2003px, Kinney:2005vj, Martin:2012pe}.
In terms of the equation of motion of the inflaton field,
\bae{
    \ddot{\varphi}+3H\dot{\varphi}+V'(\varphi)=0,
}
the friction term and the potential force are balanced with the negligible acceleration in the \ac{SR} case, while the friction dominates the potential force in the \ac{USR} case as
\bae{\label{eq: USR condition}
    \frac{\ddot{\varphi}}{H\dot{\varphi}}\simeq-3.
}
Here $\varphi$ is the homogeneous background of the inflaton field, $V$ and $V'$ are the inflaton potential and its derivative, $H$ is the Hubble parameter, and the dots represent the time derivatives.
The fact that both limiting cases lead to the scale-invariant power spectrum is understood as a consequence of the so-called Wands duality~\cite{Wands:1998yp} (see also Ref.~\cite{Tzirakis:2007bf} for a quadratic potential model) between the two independent solutions of the second-order equation of motion for the cosmological perturbation.

The scale-invariance of the power spectrum is not the unique feature of the \ac{USR} inflation.
It results in the $\calO(1)$ non-Gaussianity (in terms of the non-linearity parameter) in the primordial perturbation even from the single-field inflation~\cite{Namjoo:2012aa, Martin:2012pe, Cai:2018dkf, Passaglia:2018ixg}.
Furthermore, it has been suggested that a non-perturbative signal may arise for large perturbations as an exponentially heavy tail of the \ac{PDF} of the curvature perturbation $\calR$~\cite{Pattison:2017mbe, Cai:2018dkf, Atal:2019erb}.
According to the $\delta N$ formalism, it is understood as a result of a logarithmic map from the inflaton perturbation to the observable curvature perturbation.
Recently, in Ref.~\cite{Pi:2022ysn}, Sasaki and one of us gave a duality understanding between the two solutions of the equation of motion in this logarithmic map called the \emph{logarithmic duality}.
This clarifies the relation among the ordinary Gaussian tail in the \ac{SR} case, the exponential tail (i.e., the single exponential decay) in the \ac{USR} case, and the Gumbel tail (i.e., the double exponential decay) in certain specific examples.

In this paper, we employ the logarithmic duality shown in Ref.~\cite{Pi:2022ysn} to understand the \ac{CR} inflation~\cite{Motohashi:2014ppa, Motohashi:2017aob, Motohashi:2017vdc, Motohashi:2019tyj}, which is a generalisation of the \ac{USR} condition~\eqref{eq: USR condition}.
There, it reads a certain constant $\beta$ as
\bae{\label{eq: const roll condition}
    \frac{\ddot{\varphi}}{H\dot{\varphi}}=\beta,
}
which is not necessary $-3$.
It smoothly connects the \ac{SR} ($\beta\to0$) and \ac{USR} ($\beta\to-3$) limits as the corresponding solution~\eqref{eq: const roll condition} is an attractor for $-3/2<\beta$ while it is a non-attractor for $\beta<-3/2$.
The \ac{CR} model is also phenomenologically interesting because it can give rise to a blue-tilted spectrum to enhance the primordial perturbation on small scales and can realise exotic astrophysical objects such as primordial black holes (PBHs)~\cite{Motohashi:2019rhu, Motohashi:2023syh}.
Making use of the logarithmic duality, in this paper we show that the blue-tilted \ac{CR} attractor ($-3/2<\beta<0$) results in the Gumbel-type tail, while the red-tilted attractor ($\beta>0$) exhibits the exponential tail.

The paper is organised as follows. We first review the logarithmic duality in Sec.~\ref{sec:PiSasaki}. In Sec.~\ref{sec: const roll}, we apply the logarithmic duality to the \ac{CR} inflation and see the Gumbel tail in the \ac{PDF} of the curvature perturbation.
Sec.~\ref{sec: conclusion} is devoted to discussion and conclusions.
Throughout the paper, we adopt the Planck unit $c=\hbar=\Mpl=1$ where $\Mpl$ is the reduced Planck mass.

\section{Logarithmic duality}
\label{sec:PiSasaki}

Following Ref.~\cite{Pi:2022ysn},
we briefly review the logarithmic duality of the curvature perturbations in the single-field inflation model.
Let us use the number of $e$-folds which is defined by
\begin{equation}
	N = \int_t^{t_{\rm end}} H \dd{t} \quad (\to \dd{N} = -H \dd{t}),
\label{eq:efolds}
\end{equation}
as a time coordinate.
Assuming that the Hubble parameter, $H$, is almost constant and the potential $V(\varphi)$ of the inflaton field $\varphi$ is approximated by the quadratic form, the equation of motion for the inflaton can be considered to be~\cite{Pi:2022ysn}
\begin{equation}
	\dv[2]{\varphi}{N} -3 \dv{\varphi}{N} + 3 \eta \varphi =0,
	\label{eq:background1}
\end{equation}
where $\eta \coloneqq V''/(3 H^2)$.
The general solution is given by
\begin{align}
\varphi (N) = C_+ e^{\lambda_+ (N-N_*)} + C_- e^{\lambda_- (N-N_*)}~,  
\label{eq:gen_sol_PS}
\end{align}
with
\begin{align}
\lambda_\pm = \frac{3 \pm \sqrt{9-12 \eta}}{2}~.
\label{eq:lambda_PS}
\end{align}
Note that the two general solutions are degenerate ($\lambda_+=\lambda_-$) when $\eta=3/4$, which we do not consider. 
Then we always have $\lambda_-<\lambda_+$, and the solution with $e^{\lambda_-(N-N*)}$ is the attractor solution as it dominates in the late stage when $N\to N_*$.
Here, we assume that the phase that is controlled by the equation of motion~\eqref{eq:background1} is temporal, and denote $N_*$ as the time at the end of such a phase.
Note that we have set the potential extremum at $\varphi=0$ without loss of generality.

By taking the time derivative of this solution, we can obtain the expression for the ``field velocity", which is defined by $\pi\coloneqq-\dv*{\varphi}{N}$, as
\begin{align}
	-\pi (N) = \lambda_+ C_+ e^{\lambda_+ (N-N_*)} + \lambda_- C_- e^{\lambda_- (N-N_*)}~.      
\end{align}
Assuming $\lambda_- \neq 0$ (i.e., $\eta\neq0$), from these expressions, the number of $e$-folds is expressed in terms of $(\varphi, \pi)$ as
\begin{align}
	N - N_\ast = \frac{1}{\lambda_\pm} \ln \frac{\pi + \lambda_\mp \varphi}{\pi_\ast + \lambda_\mp \varphi_\ast} ~,  
\end{align}
where $\pi_\ast \coloneqq \pi (N_\ast)$ and $\varphi_\ast \coloneqq \varphi (N_\ast)$.
From this expression, 
the fluctuations of the number of $e$-folds from the spatially-flat hypersurface at $N$ to the uniform $\varphi_*$ hypersurface at $N_*$ is \cite{Pi:2022ysn}
\begin{align}
	\delta (N - N_\ast) = 
	\frac{1}{\lambda_\pm} \ln \left[ 1 + \frac{\delta \pi + \lambda_\mp \delta \varphi}{ \pi + \lambda_\mp \varphi} \right]
	- \frac{1}{\lambda_\pm}  \ln \left[ 1 + \frac{\delta \pi_\ast}{ \pi_\ast + \lambda_\mp \varphi_\ast} \right]~,
	\label{eq:deltaN}
\end{align}
where $\delta\varphi$ and $\delta\pi$ are the perturbation of the field and velocity on the initial hypersurface, respectively; and $\delta\pi_*\coloneqq\delta\pi(N_*)$ is the velocity perturbation on the uniform $\varphi_*$ hypersurface at $N_*$.
Note that $\pi_\ast$ is given as a function of $(\varphi, \pi)$:
\begin{align}\label{eq:conservation}
\left( \frac{\pi + \lambda_+ \varphi}{\pi_\ast + \lambda_+ \varphi_\ast} \right)^{\lambda_+} = \left( \frac{\pi  + \lambda_- \varphi}{\pi_\ast + \lambda_- \varphi_\ast} \right)^{\lambda_-}~, 
\end{align}
and $\delta \pi_\ast$ is determined by $(\delta \varphi, \delta \pi)$ as
\begin{align}
    \left( 1 + \frac{\delta \pi_\ast}{ \pi_\ast + \lambda_+ \varphi_\ast} \right)^{-\lambda_+}
    \left( 1 + \frac{\delta \pi_\ast}{ \pi_\ast + \lambda_- \varphi_\ast} \right)^{\lambda_-}
    = \left( 1 + \frac{\delta \pi + \lambda_- \delta \varphi}{ \pi + \lambda_- \varphi} \right)^{\lambda_-}
   \left( 1 + \frac{\delta \pi + \lambda_+ \delta \varphi}{ \pi + \lambda_+ \varphi} \right)^{-\lambda_+} ~. 
   \label{eq:deltapi_dual}
\end{align}
Actually, the equivalence of $\delta (N-N_\ast)$
in terms of the choice of $\lambda_\pm$ is guaranteed by the above relations. 
This equivalence is called \emph{logarithmic duality} in Ref.~\cite{Pi:2022ysn}.

\section{Constant-roll model in light of logarithmic duality}\label{sec: const roll}

Let us take a closer look at the interplay between the \ac{CR} model~\cite{Motohashi:2014ppa} and the quadratic potential.
The \ac{CR} condition is given by
\begin{align}
	\frac{\dv*[2]{\varphi}{t}}{H\dv*{\varphi}{t}} = \beta  ~,
	\label{eq:conroll}
\end{align}
where $\beta$ is a constant. It is found that the potential, in which this \ac{CR} condition is exactly satisfied, is given by~\cite{Motohashi:2014ppa}
\begin{align}
	V(\varphi) = 
	\begin{cases}
		\displaystyle 3M^2\Mpl^2\left\{1-\frac{3+\beta}{6}\left[1-\cosh\left(\sqrt{-2\beta}\frac{\varphi}{\Mpl}\right)\right] \right\}~, & (\beta < 0), \\[10pt]
        \displaystyle 3M^2\Mpl^2\left\{1-\frac{3+\beta}{6}\left[1-\cos\left(\sqrt{2\beta}\frac{\varphi}{\Mpl}\right)\right] \right\}~, & (\beta \geq 0), 
	\end{cases}
\end{align}
and the solution for the Hubble parameter is given by 
\begin{align}
	H &= 
	\begin{cases}
		M \coth (-\beta M t)~, & (\beta < 0), \\
        -M \tanh (\beta M t)~, & (\beta \geq 0).
	\end{cases}   
\end{align}

The curvature power spectrum evaluated at the horizon exit in the \ac{CR} model exhibits the scale dependence of~\cite{Motohashi:2014ppa} 
\bae{\label{eq: spectral index}
    \dv{\ln\calP_\calR(k)}{\ln k}=3-\abs{2\beta+3},
}
and hence 
the spectrum is scale-invariant for $\beta=0$ or $-3$,
red-tilted for $\beta < -3$ or $\beta > 0$, 
and blue-tilted for $-3<\beta<0$.
The red-tilted attractor $\beta>0$ can be considered as an inflationary model to generate the primordial fluctuations on \ac{CMB} scales consistent with the observational constraint~\cite{Motohashi:2014ppa, Motohashi:2017aob}.
In this case, the potential is similar to the one for the natural inflation, but with an additional negative cosmological constant. 
Therefore, for a realistic model, we have to cut the potential somewhere before it becomes negative, and it has to be changed after that in order to have subsequent reheating and radiation-dominated regimes.
Therefore, the \ac{CR} phase is temporal in this case.
On the other hand, the blue-tilted attractor $-3/2<\beta<0$ can be considered as a model to generate a large peak of the curvature power spectrum on small scales, which can lead to the formation of \acp{PBH}~\cite{Motohashi:2019rhu}.
To avoid the overproduction of \acp{PBH}, the \ac{CR} phase has to be temporal in this case as well.
Thus, in both cases, we are interested in the temporal \ac{CR} attractor phase.
Note that the \ac{CR} condition with $\beta<-3/2$ is not an attractor~\cite{Lin:2019fcz}, and the curvature perturbation grows on superhorizon scales~\cite{Motohashi:2014ppa}.

For $\varphi/\Mpl \to 0$ limit, which corresponds to $ M t \to \infty$ ($ M t \to -\infty$) for $\beta < 0$ ($\beta \geq 0$), the potential can be approximated by the quadratic potential. 
Namely, we have
\begin{align}
	V''(\varphi) \to -\beta (3+ \beta) M^2~, \qquad H \to M~,   
\end{align}
and then
\begin{align}
	\eta \coloneqq \frac{V''}{3 H^2} \to - \frac{\beta (\beta + 3)}{3}~. 
\label{eq:eta_pot}
\end{align}
Therefore, in this limit, 
we can consider the \ac{CR} model
whose dynamics is characterized by the equation of motion~\eqref{eq:background1}:
\begin{equation}
	\dv[2]{\varphi}{N} -3 \dv{\varphi}{N} + 3 \eta \varphi =0,
	\label{eq:background2}
\end{equation}
with
\begin{align}
\eta = - \frac{\beta (\beta + 3)}{3} ~, 
\label{eq:eta_constantroll}
\end{align}
which we call the \emph{quadratic constant-roll} model.
Note that, as mentioned above, we do not consider the degenerate case and hence assume $\eta\neq 3/4$, which amounts to $\beta\neq -3/2$.

Then, based on the formulation given in the previous section, we can investigate the $\delta N$ expression for the \ac{CR} model.
Note that the \ac{CR} condition \eqref{eq:conroll} with respect to the number of $e$-folds \eqref{eq:efolds} is given by
\begin{align}
    \frac{\dv*[2]{\varphi}{N}}{\dv*{\varphi}{N}} = -\beta  ~,
	\label{eq:conrollN}
\end{align}
where the Hubble parameter is approximately constant.

\subsection{Stability of constant-roll condition in quadratic model}

For the \ac{CR} inflation, there exists 
a critical value $\beta = - 3/2$, which determines 
whether the \ac{CR} condition \eqref{eq:conroll} can be satisfied as the attractor solution~\cite{Lin:2019fcz}.
Let us see how this critical value relates to the quadratic \ac{CR} model.

By substituting Eq.~\eqref{eq:eta_constantroll} into Eq.~\eqref{eq:lambda_PS}, we have
\begin{align}
\lambda_\pm = \frac{3 \pm |2 \beta + 3|}{2}~.    
\end{align}
Thus, the case with $\beta = -3/2$
gives the degenerate limit $\lambda_+ = \lambda_- = 3/2$~\cite{Pi:2022ysn}.
If $\beta > -3/2$, then we have
\begin{align}
	\lambda_+ = 3 + \beta~ \qc 
	\lambda_- = - \beta~,    
\end{align}
or if $\beta < -3/2$, then
\begin{align}
	\lambda_+ = -\beta~ \qc 
	\lambda_- = 3 + \beta~.
\end{align}
As discussed in Ref.~\cite{Pi:2022ysn}, because of $\lambda_+ > \lambda_-$, the attractor solution for \eqref{eq:gen_sol_PS} is
$\varphi (N) \propto C_- e^{\lambda_- (N-N_*)}$ which takes over at late times. 
In this sense, the \ac{CR} condition \eqref{eq:conroll} is stable for $\beta > -3/2$.
Based on the formulation discussed in \S\ref{sec:PiSasaki},
we should however take account of the other ``hidden'' decaying solution with $\lambda_+=3+\beta$ to calculate $\delta N$.

The duality between $\beta$ and $-(3+\beta)$ is nothing but the duality in \ac{CR} inflation found in Refs.~\cite{Motohashi:2014ppa, Morse:2018kda, Lin:2019fcz}.
Therefore, we expect that the logarithmic duality in the quadratic potential~\cite{Pi:2022ysn} can be generalized to the \ac{CR} model.
Below we check the above argument by solving numerically Eq.~\eqref{eq:background2}.

\subsubsection{$\beta > -3/2 $}

\begin{figure*}
     \begin{tabular}{cc}
        \begin{minipage}{0.45\hsize}
            \centering
            \includegraphics[width=7.cm]{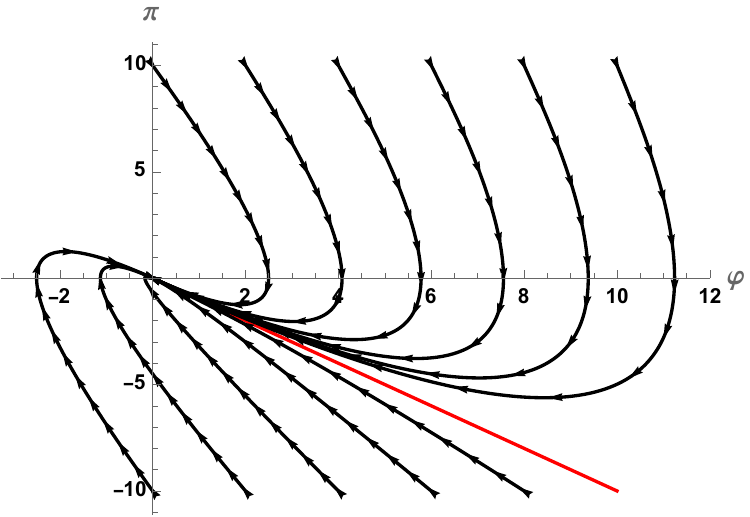}
        \end{minipage} &
        \begin{minipage}{0.45\hsize}
            \centering
            \includegraphics[width=7.cm]{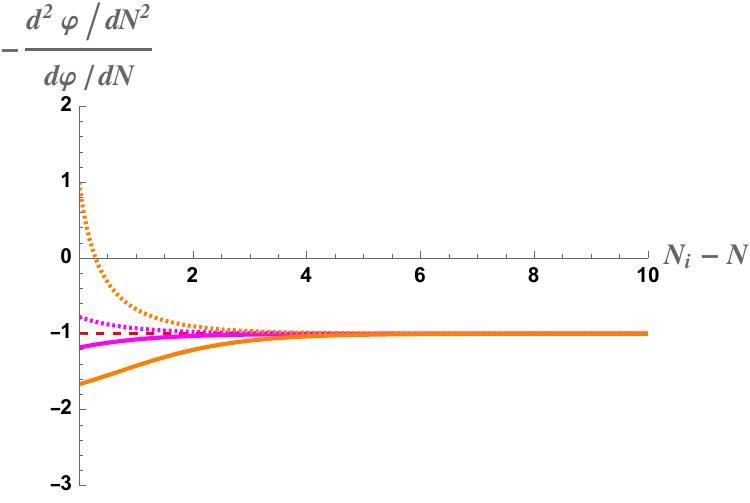}
        \end{minipage} 
    \end{tabular}
    \caption{\emph{Left}: Background trajectories in phase space for $\beta = -1$ as an example of the \ac{CR} model with $\beta > -3/2$ where the condition \eqref{eq:conroll} is stable. The red line shows the attractor solution with Eq.~\eqref{eq:attractor}.
    \emph{Right}: Attractor behaviour to the \ac{CR} phase for $\beta = -1$. The horizontal axis is the number of $e$-folds measured from the initial time, $N_i-N$, and the vertical axis is $-(\dv*[2]{\varphi}{N})/(\dv*{\varphi}{N})$. The magenta solid (dotted) line is for $\delta = 0.1$ ($-0.1$), and the orange solid (dotted) line is for $\delta = 0.5$ ($-0.5$) with $\delta$ 
    defined in Eq.~\eqref{eq:delta}. The red dashed line is the \ac{CR} condition~\eqref{eq:conrollN}.}
    \label{fig:beta_minus_1}
\end{figure*}

Based on the above discussion, in the case with $\beta > -3/2$, it is expected that the solution,
\begin{align}
\label{eq:attractor}
\varphi(N) \propto e^{-\beta (N-N_\ast)} ~\rightarrow~ \pi(N) = \beta \, \varphi(N)~,
\end{align}
is the attractor. As can be seen in the left panel of Fig.~\ref{fig:beta_minus_1}, the background trajectories indeed converge into the attractor solution at a late time.
In the right panel of Fig.~\ref{fig:beta_minus_1}, we can also check the stability with respect to the \ac{CR} condition~\eqref{eq:conrollN}.
In this plot, we change the initial condition for $\pi$ as
\begin{align}
	\pi (N_i) = \beta (1+ \delta) \varphi(N_i)~,
\label{eq:delta}
\end{align}
and vary the value of $\delta$.

\subsubsection{$\beta < -3/2 $}

\begin{figure*}
     \begin{tabular}{cc}
        \begin{minipage}{0.45\hsize}
            \centering
            \includegraphics[width=7.cm]{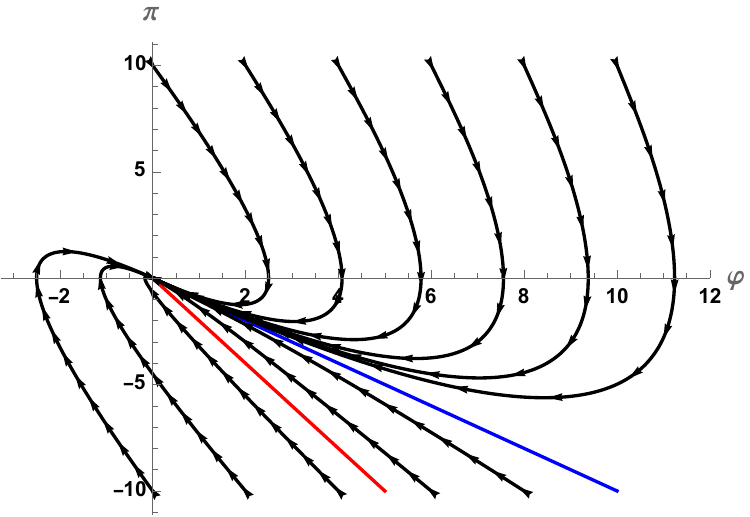}
        \end{minipage} &
        \begin{minipage}{0.45\hsize}
            \centering
            \includegraphics[width=7.cm]{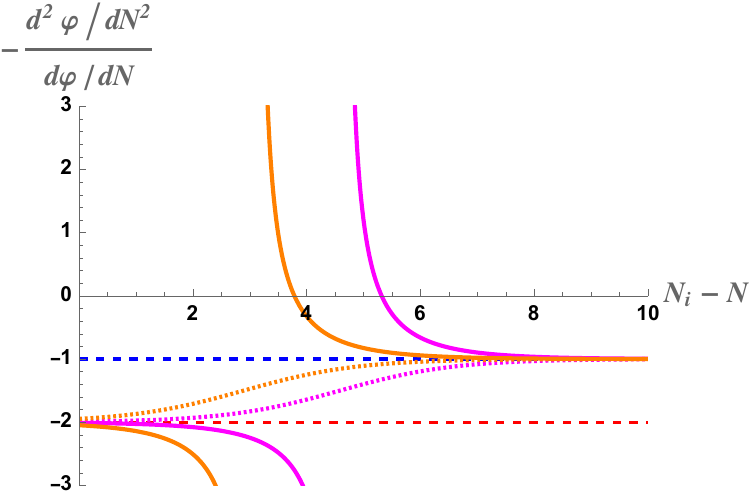}
        \end{minipage} 
    \end{tabular}
    \caption{\emph{Left}: Background trajectories in phase space for $\beta = -2$ as an example of the \ac{CR} model with $\beta < -3/2$ where the condition \eqref{eq:conroll} is unstable.
    The red line shows the solution with Eq.~\eqref{eq:attractor} and the blue one corresponds to the solution with Eq.~\eqref{eq:nonattractor2}. \emph{Right}: Attractor behaviour to the \ac{CR} phase for $\beta = -2$. The horizontal axis is the number of $e$-folds measured from the initial time, $N_i - N$, and the vertical axis is $-(\dv*[2]{\varphi}{N})/(\dv*{\varphi}{N})$. The magenta solid (dotted) line is for $\delta = 0.01$ ($-0.01$), and the orange solid (dotted) line is for $\delta = 0.05$ ($-0.05$) with $\delta$ being defined in Eq.~\eqref{eq:delta}. The red (blue) dashed line is corresponding to the solution \eqref{eq:attractor} (the solution \eqref{eq:nonattractor2}).}
    \label{fig:beta_minus_2}
\end{figure*}

\begin{figure*}
     \begin{tabular}{cc}
        \begin{minipage}{0.45\hsize}
            \centering
            \includegraphics[width=7.cm]{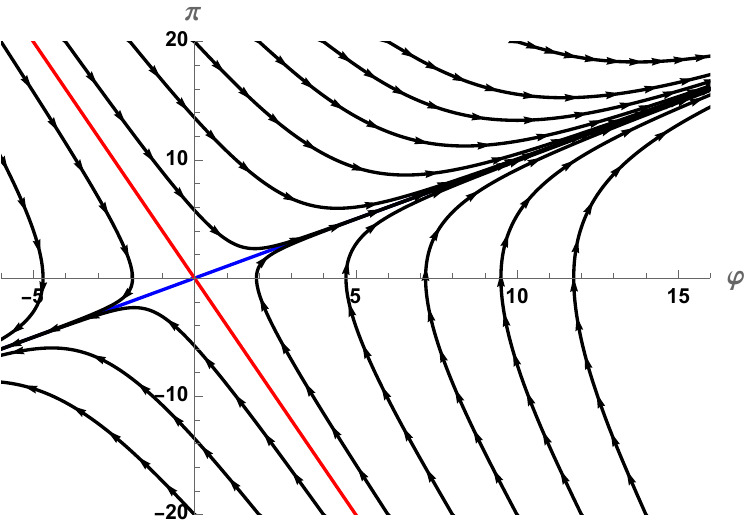}
        \end{minipage} &
        \begin{minipage}{0.45\hsize}
            \centering
            \includegraphics[width=7.cm]{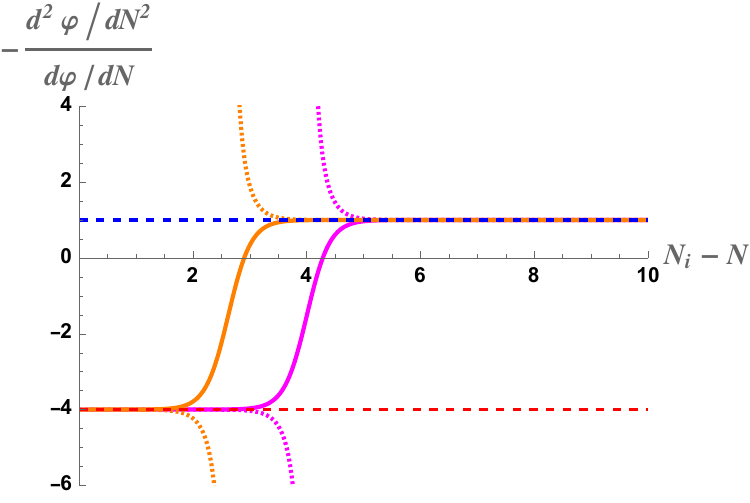}
        \end{minipage} 
    \end{tabular}
    \caption{\emph{Left}: Background trajectories in phase space for $\beta = -4<-3$. The red line shows the solution with Eq.~\eqref{eq:attractor} and the blue one corresponds to the solution with Eq.~\eqref{eq:nonattractor2}. 
    \emph{Right}: Attractor behaviour to the \ac{CR} phase for $\beta = -4$. The horizontal axis is the number of $e$-folds measured from the initial time, $N_i - N$, and the vertical axis is $-(\dv*[2]{\varphi}{N})/(\dv*{\varphi}{N})$. The magenta solid (dotted) line is for $\delta = 10^{-8}$ ($-10^{-8}$), and the orange solid (dotted) line is for $\delta = 10^{-5}$ ($-10^{-5}$) with $\delta$ being defined in Eq.~\eqref{eq:delta}. This figure is, in fact, corresponding to Figure 2 in Ref.~\cite{Pi:2022ysn}. }
    \label{fig:beta_minus_4}
\end{figure*}

Based on the above discussion, in the case with $\beta < -3/2$, it is expected that the solution~\eqref{eq:attractor} is not an attractor one, while
\begin{align}
\label{eq:nonattractor2}
\varphi (N) \propto e^{(3+\beta) (N-N_\ast)}  ~\rightarrow~ \pi (N) = -(3+\beta) \varphi (N)~,   
\end{align}
can be expected to be the attractor one, which corresponds to the $\lambda_-$-mode for the case with $\beta < -3/2$.
As shown in the left panel of Fig.~\ref{fig:beta_minus_2}, the background trajectories indeed converge into the blue line which corresponds to the solution with Eq.~\eqref{eq:nonattractor2}, not the red line which corresponds to the one with Eq.~\eqref{eq:attractor}.
In the right panel of Fig.~\ref{fig:beta_minus_2}, we also investigate the stability of the background solution for the case with $\beta = -2<-3/2$. From this figure, even for the model with constant Hubble and quadratic potential, one can also find the transition of the solution from 
$-(\dv*[2]{\varphi}{N})/(\dv*{\varphi}{N}) = \beta ~(=-2)$ to $-(3+\beta) ~(=-1)$
with the small perturbation for the initial condition, as seen in, e.g., Fig.~4 in Ref.~\cite{Lin:2019fcz}.
The similar behaviour can be checked also in the case with $\beta = -4 ~(<-3)$ as shown in Fig.~\ref{fig:beta_minus_4}.

\subsection[$\delta N$ in the constant-roll model]{\boldmath $\delta N$ in the constant-roll model}

Let us evaluate the $\delta (N-N_\ast)$ in the \ac{CR} model by making use of Eq.~\eqref{eq:deltaN}.
We assume that the inflaton undergoes a sufficiently long \ac{CR} regime, which already follows the attractor solution at $N=N_*$.
Namely, at $N=N_*$ we have 
\begin{align}\label{eq:pi=lambda_-phi}
  \pi_\ast \approx - \lambda_- \varphi_\ast,
\end{align}
where
\begin{align}
    \lambda_- = 
    \begin{cases}
        -\beta~, & (\beta > -3/2,\, \beta \neq 0), \\
        3+\beta~, & (\beta < -3/2,\, \beta \neq -3).
    \end{cases}
\end{align}
As mentioned above, we focus on the case $\lambda_- \neq 0$, i.e., we assume $\beta \neq 0$ and $\beta \neq -3$.
Either $\beta=0$ or $3$ corresponds to $\lambda_-=0$ and $\lambda_+=3$, of which the non-attractor solution is the well-known USR inflation.

Around $N=N_*$, \eqref{eq:pi=lambda_-phi} guarantees that $\pi_*+\lambda_-\varphi_*\ll \pi_*+\lambda_+\varphi_*$, which turns Eq.~\eqref{eq:deltapi_dual} into~\cite{Pi:2022ysn}
\begin{align}
    \left( 1 + \frac{\delta \pi_\ast}{ \pi_\ast + \lambda_- \varphi_\ast} \right)
    \approx \left( 1 + \frac{\delta \pi + \lambda_- \delta \varphi}{ \pi + \lambda_- \varphi} \right)
   \left( 1 + \frac{\delta \pi + \lambda_+ \delta \varphi}{ \pi + \lambda_+ \varphi} \right)^{-\frac{\lambda_+}{\lambda_-}} ~. 
\end{align}
By substituting this approximate form into Eq.~\eqref{eq:deltaN} with the upper sign,
we have
\begin{align}
\delta (N - N_\ast) &=
\frac{1}{\lambda_+} \ln \left[ 1 + \frac{\delta \pi + \lambda_- \delta \varphi}{ \pi + \lambda_- \varphi} \right]
- \frac{1}{\lambda_+}  \ln \left[ 1 + \frac{\delta \pi_\ast}{ \pi_\ast + \lambda_- \varphi_\ast} \right] \notag \\
& \approx \frac{1}{\lambda_-} \ln \left[ 1 + \frac{\delta \pi + \lambda_+ \delta \varphi}{ \pi + \lambda_+ \varphi} \right]
~.
\end{align}
This resulting expression is nothing but the first (and also dominant) term on the right-hand side of Eq.~\eqref{eq:deltaN} with the lower sign, which displays the logarithmic duality when the inflaton is already in the attractor at $N_*$.
As mentioned in Ref.~\cite{Pi:2022ysn},
in the case where the background dynamics is characterized by
Eq.~\eqref{eq:background2}, due to the linearity of the equation,
perturbations follow exactly the same equations for the background.
Therefore, $(\delta \pi + \lambda_+ \delta \varphi) / (\pi + \lambda_+ \varphi)$ should be conserved on super-horizon scales, 
and we can evaluate this at the attractor phase
where $\pi \approx - \lambda_- \varphi$ and
$\delta \pi \approx - \lambda_- \delta \varphi$
can be used. 
Finally, we can obtain the simple expression 
\begin{align}\label{eq: attractor dN}
    \delta (N - N_\ast) \simeq \frac{1}{\lambda_-} \ln \left[ 1 + \frac{\delta \varphi}{ \varphi} \right]
    = 
    \begin{cases}
      -\dfrac{1}{\beta} \ln \left[ 1 + \dfrac{\delta \varphi}{ \varphi} \right]~, & (\beta > -3/2,\, \beta \neq 0), \\[10pt]
      \dfrac{1}{3+\beta} \ln \left[ 1 + \dfrac{\delta \varphi}{ \varphi} \right]~, & (\beta < -3/2,\, \beta \neq -3).
    \end{cases}   
\end{align}

Once we assume that the curvature perturbations do not evolve after $N=N_\ast$, we can relate $\delta (N - N_\ast)$ with the late-time curvature perturbations denoted by $\mathcal{R}$.
By assuming the Gaussianity of $\delta \varphi / \varphi$, we have 
\begin{align}\label{eq:R=log}
\mathcal{R} = \delta (N-N_\ast) = 
\begin{cases}
    -\dfrac{1}{\beta} \ln \left[ 1 - \beta \, \mathcal{R}_{\rm G} \right]~, & (\beta > -3/2,\, \beta \neq 0), \\[10pt]
    \dfrac{1}{3+\beta} \ln \left[ 1 + (3+\beta) \, \mathcal{R}_{\rm G} \right]~, & (\beta < -3/2,\, \beta \neq -3).
\end{cases}   
\end{align}
where 
\begin{align}
    \mathcal{R}_{\rm G} \coloneqq 
    \begin{cases}
        -\dfrac{\delta \varphi }{\beta \varphi}~, & (\beta > -3/2,\, \beta \neq 0), \\[10pt]
        \dfrac{\delta \varphi }{(3+\beta) \varphi}~, & (\beta < -3/2,\, \beta \neq -3).
    \end{cases}
\end{align}
and then we can discuss the non-Gaussianity of $\mathcal{R}$ through this expression.

The \ac{PDF} of $\mathcal{R}$ can be obtained as
\begin{align}
    P(\mathcal{R}) &= \left| \frac{\dd \mathcal{R}_g}{\dd \mathcal{R}} \right| P_g(\mathcal{R}_g(\mathcal{R}))  \notag \\
    &= \bce{
        \displaystyle
        \frac{1}{\sqrt{2 \pi \sigma^2}} \exp \left[ - \frac{1}{2 \beta^2 \sigma^2} \left( e^{-\beta \mathcal{R}} - 1 \right)^2 - \beta \mathcal{R}\right]~, & (\beta>-3/2,\,\beta\neq0), \\[10pt]
        \displaystyle
        \frac{1}{\sqrt{2\pi\sigma^2}}\exp[-\frac{1}{2(3+\beta)^2\sigma^2}\pqty{e^{(3+\beta)\calR}-1}^2+(3+\beta)\calR], & (\beta<-3/2,\,\beta\neq-3),
    }
\end{align}
where $\sigma^2 = \langle \mathcal{R}_{\uG}^2 \rangle$.
This expression is completely equivalent to Eq.~(34) in Ref.~\cite{Pi:2022ysn}
with replacing $\lambda_-$ to $-\beta$ or $3+\beta$ for the case with $\beta > -3/2$ or $\beta < -3/2$, respectively.

Focusing on the stable \ac{CR} models with $\beta>-3/2$, we see that there are two kinds of the tail behaviour of the \ac{PDF} of the curvature perturbation $\calR$.
For the case with $-3/2 < \beta < 0$, $\lambda_- = - \beta > 0$ and this \ac{PDF} corresponds to the Gumbel-distribution-like tail as 
\bae{\label{eq:Gumbel}
    P(\mathcal{R}) \sim \exp \left[ - c^2 e^{-2 \beta \mathcal{R}} \right]~,
}
with $c^2 = (2\beta^2 \sigma^2)^{-1}$.
On the other hand, a positive $\beta$ leads to the exponential tail as
\bae{\label{eq:exp-tail}
    P(\calR)\sim\exp[-\beta \calR].
}
Note that the exponential tail has been known to show up from non-attractor models~\cite{Pattison:2017mbe, Cai:2018dkf, Atal:2019erb}.
The exponential tail~\eqref{eq:exp-tail} from attractor models is a novel example.
Recalling that the scale dependence of the power spectrum $\calP_\calR$ of the curvature perturbation is given by Eq.~\eqref{eq: spectral index}, i.e., $\dv*{\ln\calP_\calR}{\ln k}=-2\beta$ in the \ac{CR} model with $\beta>-3/2$, the tail behaviour can be classified by the spectral index: the blue-tilted curvature perturbations show the Gumbel tail while the red-tilted ones do the exponential tail.

\subsection{Consistency relation}
One may think of the so-called Maldacena's consistency relation~\cite{Maldacena:2002vr} as our \ac{CR} solution is an attractor for $\beta>-3/2$. The consistency relation claims that the correlation among $n$ hard modes and one soft mode is related to the scale dependence of the $n$-hard correlation function, known as a realisation of the cosmological soft theorems associated with the dilatation symmetry for the soft curvature perturbation or the shear transformation for the soft graviton~\cite{Assassi:2012zq, Pajer:2013ana}.
For example, the squeezed bispectrum of the curvature perturbation is related to the power spectrum by
\bae{\label{eq: consistency relation}
    B_\calR(k_1,k_2,k_3)\underset{k_1\ll k_2\sim k_3}{\sim}-\dv{\ln\calP_\calR(k_2)}{\ln k_2}P_\calR(k_1)P_\calR(k_2),
}
where
\begin{eqnarray}\label{eq: P}
    &(2\pi)^3\delta^{(3)}(\bfk_1+\bfk_2)P_\calR(k_1)=\braket{\calR_{\bfk_1}\calR_{\bfk_2}},\,
    \calP_\calR(k)=\frac{k^3}{2\pi^2}P_\calR(k), \\
    &(2\pi)^3\delta^{(3)}(\bfk_1+\bfk_2+\bfk_3)B_\calR(k_1,k_2,k_3)=\braket{\calR_{\bfk_1}\calR_{\bfk_2}\calR_{\bfk_3}},\label{eq: B}
\end{eqnarray}
with the Fourier curvature perturbation $\calR_\bfk$.
Substituting the second-order Taylor expansion of the logarithmic relation in the upper line of \eqref{eq:R=log} into the definition of the bispectrum~\eqref{eq: B} and supposing the ordinary hierarchy $P_{\calR_\uG}(k_1)\gg P_{\calR_\uG}(k_2)$ for $k_1\ll k_2$ where $P_{\calR_\uG}$ is the power spectrum for $\calR_\uG$, one obtains the squeezed bispectrum as
\bae{
    B_\calR(k_1,k_2,k_3)\underset{k_1\ll k_2\sim k_3}{\sim}2\beta P_{\calR_\uG}(k_1)P_{\calR_\uG}(k_2).
}
On the other hand, the \ac{CR} is known to exhibit the scale dependence~\eqref{eq: spectral index} 
at the leading order $\calP_\calR\simeq\calP_{\calR_\uG}$, which results in $-2\beta$ for the model with $\beta>-3/2$ where the \ac{CR} condition~\eqref{eq:conroll} can be realized as an attractor solution.
The Maldacena's consistency relation~\eqref{eq: consistency relation} is hence confirmed to hold. In other words, the non-vanishing non-Gaussianity due to the nonlinear map~\eqref{eq:R=log} can be viewed as a consequence of the scale dependence of the power spectrum from the consistency relation as inferred at the end of the previous subsection.
Note that, in contrast, the consistency relation is violated when the inflaton is in the non-attractor, such as \ac{CR} inflation with $\beta<-3/2$ including \ac{USR} inflation ($\beta=-3$)~\cite{Namjoo:2012aa,Martin:2012pe, Namjoo:2024ufv}.

\section{Discussion and conclusions}\label{sec: conclusion}

In this paper, we investigated the \ac{CR} inflation in light of the duality between solutions of the background equation of motion.
While the logarithmic duality has recently been discussed in the quadratic potential, such duality also holds in a more general \ac{CR} model.
We confirmed that the critical value $\beta\coloneqq\ddot{\varphi}/(H\dot{\varphi})=-3/2$ determining whether the \ac{CR} 
condition~\eqref{eq:conroll} is stable or not is understood as the point at which the dual solutions, i.e., the attractor and non-attractor solutions of the field equation, are interchanged.
We have shown that the blue-tilted stable \ac{CR} model, characterised by $-3/2<\beta<0$, results in the Gumbel tail in the \ac{PDF} of the curvature perturbation, $P(\calR)\sim\exp\bqty{-c^2e^{-2\beta\calR}}$, where the constant $c$ in the Gumbel tail is related to the Gaussian variance $\sigma^2=\braket{\calR_\uG^2}$ by $c^2=(2\beta^2\sigma^2)^{-1}$, while the red-tilted one, $\beta>0$, shows the exponential tail $P(\calR)\sim\exp[-\beta\calR]$.
In both cases, for $\beta > -3/2$ and $\beta \neq 0$ the curvature perturbation $\calR$ is given by a (simplified) logarithmic map of a Gaussian random field $\calR_\uG$ as
\bae{\label{eq: log map}
    \calR(\bfx)=-\frac{1}{\beta}\ln\bqty{1-\beta\calR_\uG(\bfx)}.
}
This is an interesting example of the realisations of the Gumbel tail, and also the exponential tail from an attractor model in contrast to ordinary non-attractor models exhibiting the exponential tail~\cite{Pattison:2017mbe, Cai:2018dkf, Atal:2019erb}.

In practice, the tail behavior of the curvature perturbation has a significant effect on the abundance of rare astrophysical objects such as massive galaxy clusters~\cite{Ezquiaga:2022qpw}, \acp{PBH}~\cite{Pattison:2017mbe, Atal:2019erb, Kitajima:2021fpq}, etc.
In particular, the \ac{PBH} formation requires a substantial amplification of the curvature perturbation on a smaller scale $\lesssim\si{Mpc}$, for which one needs to go beyond the simplest single-field \ac{SR} paradigm~\cite{Motohashi:2017kbs} and hence to consider, e.g., the \ac{USR} or \ac{CR} generalizations, or the multi-field extensions.
The non-Gaussian tail is a crucial feature for such cases.
It will also change the relation between the \ac{PBH} abundance and the so-called scalar-induced gravitational waves as a byproduct of the large primordial perturbation~\cite{Saito:2008jc, Saito:2009jt, Nakama:2016gzw, Garcia-Bellido:2017aan, Kohri:2018awv, Cai:2018dig, Bartolo:2018evs, Pi:2020otn, Domenech:2021ztg}.
While the impact of the exponential tail on them has been intensively studied (see, e.g., Ref.~\cite{Abe:2022xur, Pi:2024jwt}), we leave the Gumbel-tail effect for future work~\cite{Gumbel}.

\begin{figure}
    \centering
    \includegraphics[width=0.7\hsize]{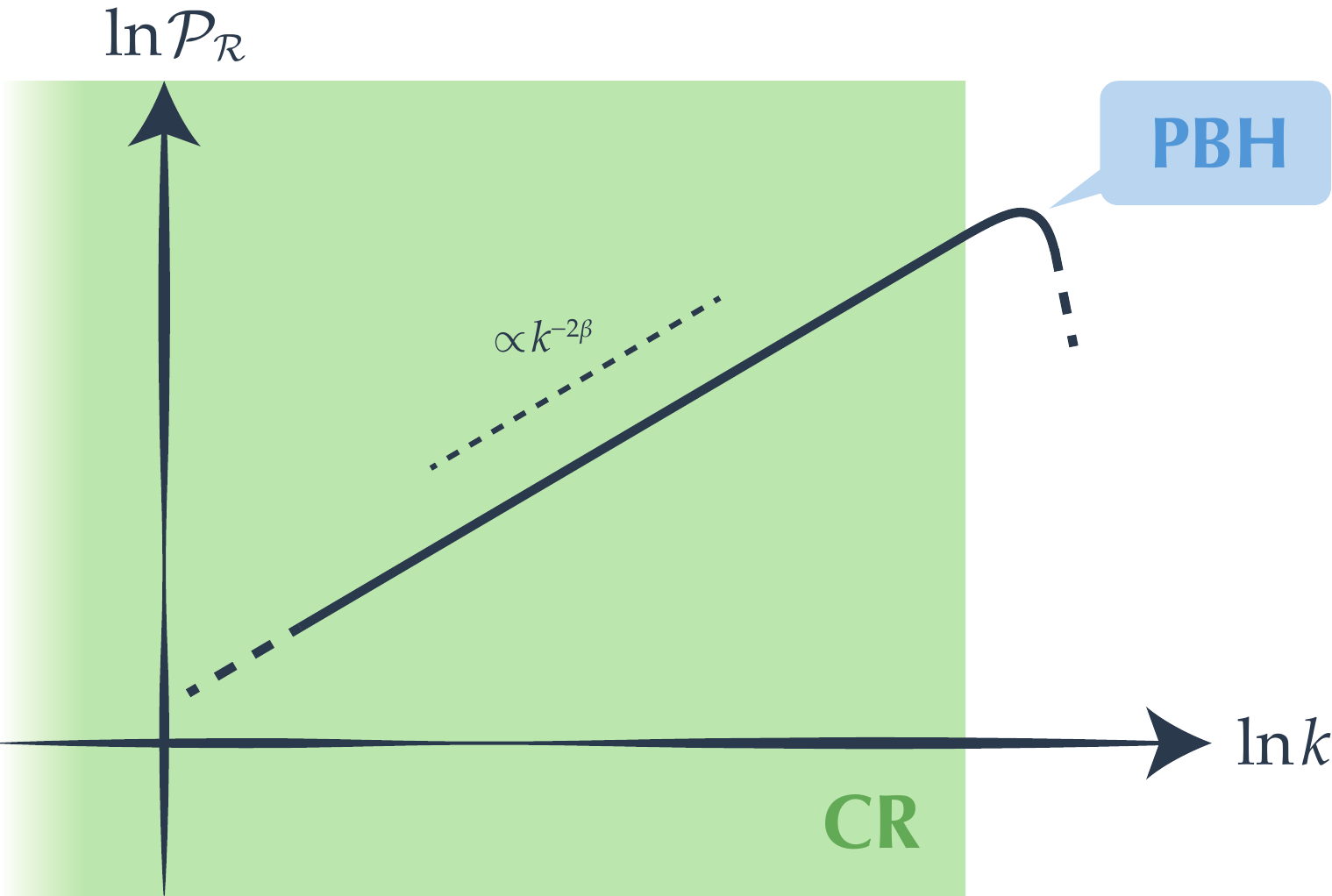}
    \caption{A schematic image of the peak of the power spectrum in a \ac{CR} model. A \ac{CR} phase with $-3/2<\beta<0$ yields a blue-tilted spectrum $\calP_\calR\propto k^{-2\beta}$ and hence the curvature perturbations are amplified toward small scales. However, it also means that the \ac{CR} phase must end for the power spectrum has a maximum $\dv*{\ln\calP_\calR}{\ln k}=0$.}
    \label{fig: CR power}
\end{figure}

We verified that Maldacena's consistency relation is valid in the attractor CR case ($\beta>-3/2$), as the quadratic expansion of \eqref{eq: log map} together with \eqref{eq: spectral index} gives
\begin{equation}
f_\mathrm{NL}=\frac56\beta=-\frac5{12}\frac{\mathrm{d}\ln\mathcal{P_R}}{\mathrm{d}\ln k}.
\end{equation}
However, 
one may note that the maximum of the power spectrum, which may correspond to the end of or after the \ac{CR} phase (see Fig.~\ref{fig: CR power} for a schematic image or Fig.~5 in Ref.~\cite{Motohashi:2019rhu} for a more realistic result), satisfies $\dv*{\ln\calP_\calR}{\ln k}=0$ by definition.
Hence, whether the Gumbel (note that the power spectrum should be blue-tilted to have a peak) or another certain non-Gaussian feature is also exhibited on the peak scale is an intriguing question, as in fact, the peak scale will be the most relevant to the \ac{PBH} formation.
To clarify this point, it is necessary to consider not a simple setup such as the one considered here, but a specific model that can realize the end of the inflation, for example by connecting the \ac{CR} or \ac{USR} phase to a standard slow-roll one, e.g., as discussed in Refs.~\cite{Garcia-Bellido:2017mdw, Motohashi:2017kbs, Motohashi:2019rhu, Motohashi:2023syh}.
We also leave it for future work.

\section*{Note added}

During the preparation of our paper,~\cite{Wang:2024xdl} appeared in arXiv, which obtained a similar duality understanding of the \ac{CR} inflation. The explicit illustration of the Gumbel tail is our novelty.

\acknowledgments
We would like to thank Qing Gao and Yungui Gong for discussions and useful comments. 
This work is supported in part by the National Key Research and Development Program of China Grant No.~2021YFC2203004 (SP),
and by JSPS KAKENHI Grants
No.~JP22K03639 (HM), JP24K00624 (SP), 
JP21K13918 (YT), JP24K07047 (YT),
JP20K03968 (SY), JP24K00627 (SY), and JP23H00108 (SY). 
SP is also supported by National Natural Science Foundation of China No. 12475066 and No.12047503. 
SP and SY are also supported by the World Premier International Research Center Initiative (WPI Initiative), MEXT, Japan.
RI is supported by JST SPRING, Grant Number JPMJSP2125, and the ``Interdisciplinary Frontier Next-Generation Researcher Program of the Tokai Higher Education and Research System''. 

\bibliographystyle{JHEPmod}
\bibliography{ref}

\end{document}